\documentclass[preprint,showpacs,nofootinbib,prd,aps]{revtex4-1}
\usepackage{graphicx,amstext,amssymb}

\begin{document}
\title{ Initialization of hydrodynamics in relativistic heavy ion
collisions with an energy-momentum transport model}

\author{V.Yu. Naboka$^1$}
\author{S.V. Akkelin$^{1,2}$}

\author{Iu.A. Karpenko$^{1,3}$}
\author{Yu.M. Sinyukov$^1$}
\affiliation{$^1$Bogolyubov Institute for Theoretical Physics,
Metrolohichna  14b, 03680 Kiev,  Ukraine} \affiliation{$^2$Institut
f\"ur Theoretische Physik, Universit\"at Heidelberg, Philosophenweg
16, 69120 Heidelberg, Germany} \affiliation{$^3$ Frankfurt Institute
for Advanced Studies, Ruth-Moufang-Stra\ss e 1, 60438 Frankfurt am
Main, Germany}

\begin{abstract}

A key ingredient of hydrodynamical modeling of relativistic heavy
ion collisions is thermal initial conditions,  an input that is the
consequence of a pre-thermal dynamics which is not completely
understood yet.  In the paper we employ a recently developed
energy-momentum transport  model of the pre-thermal stage to study
influence of the alternative  initial states in nucleus-nucleus
collisions on flow and energy density distributions of the matter at
the starting time of hydrodynamics.  In particular, the dependence
of the results on isotropic and anisotropic initial states is
analyzed. It is found that at the thermalization time the transverse
flow is larger and the maximal energy density is higher for the
longitudinally squeezed initial momentum distributions. The results
are also sensitive to the relaxation time parameter, equation of
state at the thermalization time, and transverse profile of initial
energy density distribution: Gaussian approximation, Glauber Monte
Carlo profiles, etc. Also, test results ensure that the numerical
code based on the energy-momentum transport model is capable of
providing both averaged and fluctuating initial conditions for the
hydrodynamic simulations of relativistic nuclear collisions.

\end{abstract}

\pacs{25.75.-q,  24.10.Nz}

 \maketitle

 \section{Introduction}

Hydrodynamics is considered  now as an integral part of a future
``Standard Model'' for the evolution of the Little Bang fireballs
created in relativistic heavy ion collisions at the Relativistic
Heavy Ion Collider (RHIC) and the Large Hadron Collider (LHC) (for
up-to-date reviews, see Ref. \cite{Song}). To complete the
development of the ``Standard Model'', a hydrodynamical approach
must be supplied with initialization and breakup conditions: The
former ones should describe transition from a dense non-equilibrated
state to a near local equilibrium one, and the later ones form a
prescription for particle production during the breakup of the
continuous medium at the final stage of hydrodynamical expansion.

Until now the main progress  was reached in understanding and
modeling  the breakup conditions at the later dilute stage of matter
expansion when the hydrodynamical approximation is no longer valid.
Namely, it is widely accepted that a quark-gluon fluid is followed
by the hadronic gas that is highly dissipative and evolves away from
equilibrium. The transition between the quark-gluon fluid and
hadronic gas is described by means of the so-called hybrid models
where conversion of the fluid to particles is typically realized at
a hypersurface of hadronization or chemical freeze-out\footnote{ It
has been well known for a long time that such a matching
prescription has problems with the energy-momentum conservation laws
when fluid is converted to particles at a hypersurface which
contains non-space-like parts. These problems can be avoided by
using the hydrokinetic approach that was proposed in Ref.
\cite{hydrokin-1} and further developed in Ref. \cite{hydrokin-2}
(see also Ref. \cite{hydrokin-3}), which accounts for continuous
particle emissions   during the whole period of hydrodynamic
evolution and is based not on the distribution function  but on the
escape one.} by a Monte Carlo event generator (for recent
discussions of the particlization procedure see, e.g., Refs.
\cite{hydrokin-3,convert}), and subsequent hadronic stage of
evolution  is modeled by a hadronic cascade model like UrQMD
\cite{urqmd}.

As for initialization of the hydrodynamical evolution, one needs to
note that presently  there is no commonly accepted model of the
pre-equilibrium dynamics and subsequent thermalization  (for the
discussions of possible mechanisms of thermalization see, e.g., Ref.
\cite{Mull}).   There is, however,  theoretical  evidence \cite{CGC}
that the state which emerges in relativistic heavy ion collisions
possesses large momentum-space anisotropies in the local rest
frames. Such  an initial state is far from equilibrium and can not
be utilized as an input for hydrodynamics. Because  initial state
fluctuates on an event-by-event basis, Monte Carlo event generators
are widely used for the generation of the initial states  in
relativistic $A+A$ collisions. The models of initial state most
commonly used now are MC-Glauber (Monte Carlo Glauber) \cite{MCG},
MC-KLN (Monte Carlo Kharzeev-Levin-Nardi) \cite{MC-KLN}, and
IP-Glasma (impact-parameter-dependent Glasma) \cite{IPG}. The latter
model also includes some non-equilibrium dynamics of the gluon
fields which, however, does not lead to a proper equilibration. To
apply these models for data description, some thermalization process
has to be assumed. Evidently, in order to reduce uncertainties of
results obtained by means of hydrodynamical models, one needs to
evolve far-from-equilibrium initial state of matter in
nucleus-nucleus collision to a close to locally equilibrated one by
means of a reasonable pre-equilibrium dynamics.

It is well known that for far-from-equilibrium systems  one can not
use the Gibbs thermodynamic relations to get the equations
expressing the conservation laws in the system in the closed form as
 is done in hydrodynamics. The latter is, in fact, an effective
theory which describes long wavelength dynamics of systems that are
close to (local) equilibrium (see, e.g., Ref. \cite{hyd} and
references therein). As for far-from-equilibrium systems, the
underlying kinetics has to be used in direct form to enclose  the
energy-momentum balance equations. Typically,  even if  underlying
kinetic equations are known, the (approximate) solution of these
equations is known only in the vicinity of the (local) equilibrium
state of a system. For example, kinetic derivation of the viscous
hydrodynamical equations for dilute gases is based on approximate
solutions of the Boltzmann equations near the local equilibrium
distribution (see, e.g., Ref. \cite{Huang}). Sometimes, if proper
kinetics is unknown or too complicated, the relaxation time
approximation of the collision term is utilized (see Ref. \cite{RTA}
for the relativistic case). Depending on a value of the time-scale
relaxation parameter, a solution of such a kinetic equation
interpolates between the two trivial limiting cases: free streaming
and locally equilibrated evolutions. Although the relaxation time
approximation has been  known for a long time, it is used relatively
rarely for practical calculations of far-from-equilibrium dynamics
because the corresponding kinetic equation has to be accompanied by
the conservation law constraints for the collision term (e.g.,
Landau matching conditions), which  result in nonlinear equations.
This  is the reason why finding solutions of kinetic equations in
the relaxation time approximation typically requires time-consuming
numerical calculations (especially for $3$ dimensional dynamics).

Recently, an approach  called ``anisotropic hydrodynamics'' was
developed to account for large early-time deviations from local
equilibrium  in relativistic heavy ion collisions in a
hydrodynamic-like manner (for review, see Ref. \cite{aniz} and
references therein). The zeroth and the first moments of the $0+1$
kinetic equation in the relaxation time approximation were  used to
find the evolutionary equations for the parameters of the
boost-invariant Romatschke-Strickland form \cite{RSF} of the
one-particle distribution function with the help of the Landau
matching conditions and  exponential Romatschke-Strickland ansatz
\cite{Mart-1}. Then, utilization of the Romatschke-Strickland
distribution function allows one  to calculate  the non-equilibrium
energy-momentum tensor and express thermodynamic-like quantities
(which do not have standard thermodynamic interpretation) as
functions of some parameters in an equation-of-state manner, closing
in such a way the system of the energy-momentum conservation
equations. Despite the fact that the Romatschke-Strickland
distribution function does not satisfy the kinetic equation but some
moments only, it was demonstrated that the energy-momentum tensor of
anisotropic hydrodynamics approximates well the far-from-equilibrium
energy-momentum tensor that is calculated from exact numerical
solution of $0+1$ kinetic equation in the relaxation time
approximation for a system which is transversely homogeneous and
undergoing boost-invariant longitudinal expansion \cite{Mart-2}.

Very recently, various attempts were performed  to generalize the
anisotropic hydrodynamics framework  to describe
far-from-equilibrium dynamics beyond the $0+1$ dimensions, see,
e.g., Ref. \cite{Mart-3}. Unlike $0+1$ dimensional case, such
generalizations were not compared with exact kinetics, and their
relevance for description of far-from-equilibrium dynamics remains
questionable. In particular, to justify utilization of the
generalized ``equations of state'' based on  the
Romatschke-Strickland ansatz beyond the $0+1$ dimensions, the
concept of the ``anisotropic equilibrium'' has been introduced. The
problem is that the far-from-equilibrium Romatschke-Strickland
ansatz of the distribution function does not solve the corresponding
kinetic equation, even approximately, and utilization of such an
ansatz as ``leading order'' approximation does not have solid
ground. It is different from  the standard second-order
(Israel-Stewart) viscous hydrodynamics, where expansion around local
equilibrium distribution is justified in the vicinity of a high
entropy local equilibrium state. As we noted above, this ansatz and
the corresponding ``equations of state'' are grounded, in fact,  on
the boost-invariant transversely homogeneous kinetics, and therefore
can hardly  provide an adequate approximation of non-trivial
transverse dynamics. It especially concerns calculations on an
event-by-event basis, where a typical initial state is  highly
inhomogeneous in the transverse plane and produces locally large
transverse velocities. On the other hand, utilization of
transversely homogeneous initial conditions with very specific type
of the initial anisotropy seriously restricts the scope of
applicability of the anisotropic hydrodynamics.

In this article we use another  phenomenological  approach, proposed
in Ref. \cite{preth}. This approach  allows pre-equilibrium dynamics
to be matched to a hydrodynamic description. The method is based on
the energy-momentum conservation equations that are associated with
the relaxation transport dynamics, expressed for energy-momentum
tensor that evolves towards its hydrodynamical form. It allows one,
using the relaxation time parameter, to assess the hydrodynamic
energy-momentum tensor at the assumed time of thermalization
starting from any initial one. The key feature of the method is that
there are no additional assumptions (such as, e.g., the Landau
matching conditions or ``anisotropic equilibrium'' concept) needed
to describe the transition from the far-from-equilibrium regime to
the  near local equilibrium one. Then this model can continuously
interpolate between  a far-from-equilibrium initial state, with
arbitrary type of anisotropy, and the regime described by the
hydrodynamics. Moreover, the method allows one to account for large
initial state inhomogeneities that lead to non-trivial transverse
dynamics. Therefore the method may be used to model the very early
stages of relativistic heavy ion collisions on an event-by-event
basis. Here we develop a numerical realization of this method,
aiming to study the connection between initial locally isotropic or
anisotropic momentum space distributions, and the equilibrium
initial conditions for subsequent hydrodynamical evolution in
relativistic nuclear collisions. In this article we restrict
ourselves to the central rapidity region, where the longitudinal
boost invariance seems to be a good enough approximation to the
longitudinal dynamics.

\section{Energy-momentum  relaxation dynamics for a far-from-equilibrium initial state}

It was proposed in Ref. \cite{preth} to simulate the approach to local
equilibrium  of the matter produced in ultrarelativistic heavy ion
collisions by means  of the relaxation dynamics  of the
energy-momentum tensor which is motivated by  Boltzmann kinetics in
the relaxation-time approximation,
\begin{eqnarray}
 \frac{p^{\mu} \partial f(x,p)}{p^0\partial x^{\mu }}
 =-\frac{f(x,p)-f_{\text{l
 eq}}(x,p)}{\tau_{\text{rel}}}, \label{boltz-1}
\end{eqnarray}
where $\tau_{\text{rel}}$ is the relaxation time parameter in the
center of  mass reference frame (in general it can be  some function
of $(x,p)$), $f(x,p)$ and $f_{\text{l
 eq}}(x,p)$ are actual  and local-equilibrium
phase-space distribution functions, respectively, and the energy-momentum
tensor is defined as
\begin{eqnarray}
T^{\mu \nu}(x)=\int d^{3}p \frac{p^{\mu}p^{\nu}}{p_{0}}f(x,p).
\label{002}
\end{eqnarray}
As one can see from Eq. (\ref{boltz-1}), the target
(local-equilibrium) state is reached in a finite time interval at $t
= t_{\text{th}}$ only if the relaxation time parameter in Eq.
(\ref{boltz-1}) vanishes at $t\rightarrow t_{\text{th}}$:
$\tau_{\text{rel}}(t\rightarrow t_{\text{th}},\textbf{r},p)
\rightarrow 0$.

In the relaxation-time approximation of kinetics, the actual
distribution function, $f(x,p)$, is functional of the (target) local
equilibrium distribution function, $f_{\text{l
 eq}}(x,p)$. The formal solution of Eq. (\ref{boltz-1}) reads
\begin{eqnarray}
f(t,{\bf r},p) =f(t_{0},{\bf r}-\frac{{\bf p}}{p_{0}}
(t-t_{0}),p)P(t_{0},t,\textbf{r},p) + \nonumber \\
\stackrel{t}{%
\mathrel{\mathop{\int }\limits_{t_{0}}}%
}f_{\text{l eq}}(t^{\prime },{\bf r}-\frac{{\bf
p}}{p_{0}}(t-t^{\prime }),p) \frac{d}{dt^{\prime }}P(t^{\prime}, t, \textbf{r},p) dt^{\prime},
 \label{boltz-3-1}
\end{eqnarray}
where
\begin{equation} P(t^{\prime}, t,\textbf{r},p)=
\exp  \left\{{-\stackrel{t}{%
\mathrel{\mathop{\int }\limits_{t^{\prime }}}%
}\tau^{-1}_{\text{rel}}(s,{\bf r}-\frac{{\bf
p}}{p_{0}}(t-s),p)}ds\right\} \label{boltz-3-2}
\end{equation}
 is the probability for the particle with momentum $\textbf{p}$
to propagate freely from point $(t^{\prime},{\bf r}-\frac{{\bf
p}}{p_{0}}(t-t^{\prime}))$ to point $(t,{\bf r})$, and $f(t_{0},{\bf
r}-\frac{{\bf p}}{p_{0}} (t-t_{0}),p)\equiv f_{\text{free}}(t,{\bf r},p)$
is free streaming initial
distribution: $p^{\mu} \partial_{\mu} f(t_{0},{\bf r}-\frac{{\bf
p}}{p_{0}} (t-t_{0}),p)=0$.

Computational complexity of finding the local equilibrium state
parameters makes an utilization of Eq. (\ref{boltz-1}) or
its formal solution (\ref{boltz-3-1}) difficult for a matching of a
far-from-equilibrium initial state with perfect or viscous
hydrodynamics in relativistic heavy ion collisions. To make the
problem tractable, it was proposed in Ref.  \cite{preth} to utilize
the relaxation dynamics of the energy-momentum tensor  that
approximates the most important properties of the formal solution
(\ref{boltz-3-1}) of the  relaxation-time kinetics (\ref{boltz-1})
and, simultaneously, allows one to avoid computational problems
related to  nonlinear equations for the parameters of the local
equilibrium distribution.

For the reader's convenience, in this Section we briefly summarize
the main features of the relaxation dynamics of the energy-momentum
tensor relevant to our work, referring the reader to Ref.
\cite{preth} for more details. First, note that the boost-invariant
scenario in central rapidity region   with Bjorken longitudinal
proper time $\tau = \sqrt{t^{2}-z^{2}}$ is used in this model, and
particle probability to fly freely from the initial time $\tau_0$ to
time $\tau$ is taken in the form: $P(\tau_{0}, \tau, {\bf r},p)
\approx P(\tau_{0}, \tau)\equiv P(\tau)$. Then  the phase-space
distribution function  $f_{\text{free}}P +f_{\text{l eq}}(1-P)$ is a
formal solution of Eq. (\ref{boltz-1}) if the term
$(1-P(\tau))p^{\mu}\partial_{\mu} f^{\text{l eq}}(x,p)$ is
neglected. Correspondingly, the non-equilibrium energy-momentum
tensor  reads
\begin{eqnarray}
T^{\mu \nu}(x)=T^{\mu \nu}_{\text{free}}(x){\cal
P}(\tau)+T_{\text{hyd}}^{\mu \nu}(x)(1-{\cal P}(\tau)), \label{1}
\end{eqnarray}
where $T^{\mu \nu}_{\text{free}}(x)$ and $T_{\text{hyd}}^{\mu
\nu}(x)$ are the energy-momentum tensors of the free streaming and
hydrodynamical (local equilibrium)  components, respectively.  We
use in Eq. (\ref{1}) the substitution $P(\tau) \rightarrow {\cal
P}(\tau)$ because further we consider $0\leq {\cal P}(\tau) \leq 1$
just as an interpolating function,  and approximation (\ref{1}) will
be applied for any kind of systems, not only  for Boltzmann gas,
with target energy-momentum tensor corresponding to relativistic
ideal as well as  viscous fluids. One can see from Eq. (\ref{1})
that the following equalities
 have to be satisfied:
\begin{eqnarray}
{\cal P}(\tau_0)=1, \quad  {\cal P}(\tau_{\text{th}})=0, \quad
\partial_{\mu}{\cal P}(\tau)|_{\tau=\tau_{\text{th}}}=0.
 \label{3}
\end{eqnarray}
For interpolation function, ${\cal
P}(\tau)$, we use an ansatz proposed in Ref. \cite{preth}:
\begin{eqnarray}
{\cal P}(\tau)=  \left (
\frac{\tau_{\text{th}}-\tau}{\tau_{\text{th}}-\tau_{0}}\right
)^{\frac{\tau_{\text{th}}-\tau_0}
 {\tau_{\text{rel}} }}.
 \label{2}
\end{eqnarray}
Here $\tau_0$ is the time when relaxation dynamics is started, and
it can be  chosen as close as possible to the time when the nuclear
overlap is completed and initial non-equilibrated superdense state
of matter is formed.  The time-scale parameter $\tau_{\text{rel}}$
regulates steepness of the transition to hydrodynamics, and
self-consistency of the model, Eq. (\ref{3}), requires
$\frac{\tau_{\text{th}}-\tau_0} {\tau_{\text{rel}}}>1$ \cite{preth},
that is a constraint on the model parameters. We choose
$\tau_{0}=0.1$ fm/c and keep this parameter to be fixed throughout
all the model calculations, as well as $\tau_{\text{th}}=1$ fm/c
which is assumed to be the time when transition to hydrodynamics is
fulfilled.\footnote{ In hydrodynamic models  which ignore the
pre-equilibrium dynamics a very early initial time around $0.4-0.6$
fm/c is typically utilized to develop fairly strong transverse flows
and so describe the data in heavy ion collisions at RHIC and LHC
energies. Such a fast isotropization and thermalization of the
system is rather questionable. On the other hand,   the
pre-equilibrium dynamics of the  system generates flow at early
times, and allows one to start  hydrodynamics  with initial flows at
later times, see Ref. \cite{early}.}

To specify  the energy-momentum tensor of the free streaming
component  in Eq. (\ref{1}), notice that initially
$T_{\text{free}}^{\mu \nu}(x)$ coincides with $T^{\mu \nu}(x)$, see
Eqs. (\ref{1}) and (\ref{3}). Then the initial conditions for the former
are the same as for the latter and thus are defined by an initial
state of matter in a nucleus-nucleus collision. Further evolution of
$T_{\text{free}}^{\mu \nu}(x)$  depends  on the type of the system
and its evolution with almost no interactions. In the
further calculations we assume that such a dynamics is governed by
the one-particle  distribution function for scalar massless particles
(partons), $f(x,p)$, which satisfies the free evolution equation
\begin{eqnarray}
p_{\mu}\partial^{\mu} f(x,p)=0. \label{4}
\end{eqnarray}
The  corresponding energy-momentum tensor, $T_{\text{free}}^{\mu
\nu}(x)$,  is then  evaluated from Eq. (\ref{002}).

The energy-momentum tensor of the hydrodynamical component,
$T_{\text{hyd}}^{\mu \nu}(x)$,  is taken in its familiar form,
\begin{eqnarray}
T^{\mu \nu}_{\text{hyd}}(x)= (\epsilon_{\text{hyd}}(x) +
p_{\text{hyd}}(x)+\Pi)u^{\mu}_{\text{hyd}}(x)u^{\nu}_{\text{hyd}}(x)
- (p_{\text{hyd}}(x)+\Pi)g^{\mu \nu} +\pi^{\mu \nu}.
 \label{6}
\end{eqnarray}
Here, $u^{\mu}$  is the four-vector energy flow field,
$\epsilon_{\text{hyd}}$ is energy density in the fluid rest frame,
$p_{\text{hyd}}$ is equilibrium pressure, $\pi^{\mu \nu}$ is the
shear stress tensor, and $\Pi$ is the bulk pressure. In the present
paper we neglect bulk pressure,  and  for the shear stress tensor
use the equation of motion as in Ref. \cite{Karp},
\begin{eqnarray}
\langle u^\gamma \partial_{;\gamma} \pi^{\mu\nu}\rangle
=-\frac{\pi^{\mu\nu}-\pi_\text{NS}^{\mu\nu}}{\tau_\pi}-\frac 4 3
\pi^{\mu\nu}\partial_{;\gamma}u^\gamma, \label{evolution-1}
\end{eqnarray}
where $\partial_{;\mu}$ denotes a covariant derivative (see, e.g.,
Ref. \cite{Karp}), brackets in Eq. (\ref{evolution-1}) are defined
as: $\langle A^{\mu\nu}\rangle=(\frac 1 2 \Delta^\mu_\alpha
\Delta^\nu_\beta+\frac 1 2 \Delta^\nu_\alpha \Delta^\mu_\beta -
\frac 1 3 \Delta^{\mu\nu}\Delta_{\alpha\beta})A^{\alpha\beta}$,
$\Delta^{\mu\nu}=g^{\mu\nu}-u^\mu u^\nu$, and
$\pi^{\mu\nu}_\text{NS}$ is the values of shear stress tensor in
limiting Navier-Stokes case,
\begin{eqnarray}
\pi^{\mu\nu}_\text{NS}&=\eta(\Delta^{\mu\lambda}\partial_{;\lambda}u^\nu+\Delta^{\nu\lambda}\partial_{;\lambda}u^\mu)-\frac
2 3 \eta\Delta^{\mu\nu}\partial_{;\lambda}u^\lambda.
\label{evolution-2}
\end{eqnarray}

The evolutionary equations for  $T_{\text{hyd}}^{\mu \nu}(x)$ follow
from the  energy-momentum conservation laws, $\partial_{;
\mu}T^{\mu \nu}(x)=0$. They are
\begin{eqnarray}
\partial_{;\mu}[(1-{\cal P}(\tau))T^{\mu
\nu}_{\text{hyd}}(x)]= - \partial_{;\mu}[T^{\mu
\nu}_{\text{free}}(x) {\cal P}(\tau)]. \label{7}
\end{eqnarray}
Now, let us  take into account that $T_{\text{free}}^{\mu \nu}(x)$
is subjected to the free streaming dynamics, and
$\partial_{;\mu}T^{\mu \nu}_{\text{free}}(x)=0$. Also, let us
introduce the tensor $\widetilde{T}^{\mu \nu}_{\text{hyd}}(x)$ that
is the re-scaled hydrodynamic tensor, $\widetilde{T}^{\mu
\nu}_{\text{hyd}}(x)=(1-{\cal P}(\tau))T^{\mu \nu}_{\text{hyd}}(x)$,
with initial conditions $\widetilde{T}^{\mu \nu}_{\text{hyd}}(x)=0$
at $\tau = \tau_{0}$ everywhere in space. Then Eq. (\ref{7}) takes
its final form, the form of the hydrodynamical equation with the
source term on the right-hand side:
\begin{eqnarray}
\partial_{;\mu}\widetilde{T}^{\mu
\nu}_{\text{hyd}}(x)= - T^{\mu \nu}_{\text{free}}(x)\partial_{;\mu}
{\cal P}(\tau). \label{8}
\end{eqnarray}
By multiplying Eq. (\ref{evolution-1}) by $(1-{\cal P})$ and
substituting $\pi^{\mu\nu} =\widetilde{\pi}^{\mu\nu}/(1-{\cal P})$,
we get the equation for the re-scaled shear stress tensor
$\widetilde{\pi}^{\mu\nu}$:
\begin{eqnarray}
(1-{\cal P}(\tau))\left \langle u^\gamma \partial_{;\gamma}
\frac{\widetilde{\pi}^{\mu\nu}}{(1-{\cal P}(\tau))}\right \rangle
=-\frac{\widetilde{\pi}^{\mu\nu}-(1-{\cal
P}(\tau))\pi_\text{NS}^{\mu\nu}}{\tau_\pi}-\frac {4}{3}
\widetilde{\pi}^{\mu\nu}\partial_{;\gamma}u^\gamma.
\label{evolution-3}
\end{eqnarray}
In what follows, we take into account that the net baryon density is
small at the top RHIC and the LHC energies, and therefore neglect its
influence on the equation of state (EoS), etc.

To close the set of evolutionary equations (\ref{8}) one needs to
specify EoS $p_{\text{hyd}}=p_{\text{hyd}}(\epsilon_{\text{hyd}})$
in the hydrodynamic component.  If it is done, then Eq. (\ref{8})
allows one to deduce the initial conditions for subsequent
hydrodynamical evolution by evolving $\widetilde{T}^{\mu
\nu}_{\text{hyd}}(x)$. It is so because  the source term in Eq.
(\ref{8}) finally (at $\tau = \tau_{\text{th}}$) disappears, and
$\widetilde{T}^{\mu \nu}_{\text{hyd}}(x) \rightarrow T^{\mu
\nu}_{\text{hyd}}(x)\rightarrow T^{\mu \nu}(x)$ when $\tau
\rightarrow \tau_{\text{th}}$,  see Eq. (\ref{3}). In the next
Section we present and discuss the results of numerical implementation
of the early-stage relaxation model  for initialization of
hydrodynamics.

\section{Results and discussion}

To perform simulations of the pre-equilibrium  dynamics within the
model, we modify the code described in Ref. \cite{Karp} to solve
hydrodynamical equations with extra source terms. To make
calculations less time consuming, we reduce the original $3+1$
dimensional viscous hydrodynamic code to a $2+1$ dimensional case
assuming longitudinal boost invariance. Also, because it is known
that the viscosity-to-entropy ratio of the quark-gluon fluid is near
its minimal value at RHIC and LHC energies,  we perform some of the
simulations in the limit of zero viscosity aiming to reveal
principal features of the relaxation dynamics. We  perform numerical
calculations for the  EoS in its simplest form, $p_{\text{hyd}}=
\textit{const} \cdot \epsilon_{\text{hyd}}$. We set
$\tau_{\text{rel}}=0.5$ fm/c as a default value.

To initialize the simulations,  one needs to specify the initial
conditions. Because $\widetilde{T}^{\mu \nu}_{\text{hyd}}(x)$ is
equal to zero initially,  the corresponding initial conditions are
determined by the explicit form of the source term in the right-hand
side of Eq. (\ref{8}). Inasmuch as ${\cal P}(\tau)$ is explicitly
defined, see Eq. (\ref{2}), it remains to define the initial value of
the energy-momentum tensor of the free streaming component. We
define it by means of Eq. (\ref{002}) through initial value of the
phase space density $f(x,p)$.  In
what follows, for aim of comparisons of our calculations, we
normalize all initial distributions  in such a way that  the energy
density in the center of the system (which coincides with the center
of coordinates) is equal to $1000$ GeV fm$^{-3}$. It is a typical
value  for the simulations of $Pb+Pb$ collisions in hydrokinetic model
 with  initial time $0.1$ fm/c \cite{hydrokin-2}.

We employ here the
analytical parametrization of the initial phase-space density taken from Ref. \cite{parametr-2}.
This longitudinally boost-invariant parametrization  allows one to account for
anisotropy in momentum space. We assume no transverse flow at the
initial time $\tau_{0}=0.1$ fm/c, thereby  the initial phase-space
density does not have $x-p$ correlations in the transverse plane.
Also,  we supplement the momentum distribution from Ref.
\cite{parametr-2} with the Gaussian spatial distribution,
$\rho(\textbf{r}_T)$:
\begin{eqnarray}
\rho(\textbf{r}_T)=\exp(-r_x^2/R_x^2-r_y^2/R_y^2). \label{9.1}
\end{eqnarray}
Then  at the initial time $\tau_{0}=0.1$ fm/c
\begin{eqnarray}
f(x,p)=g \exp\left(-\sqrt{\frac{(p\cdot U)^2-(p\cdot
V)^2}{\lambda_{\perp}^2}+\frac{(p\cdot
V)^2}{\lambda_{\parallel}^2}}\right)\rho(\textbf{r}_{T}). \label{10}
\end{eqnarray}
Here $ \eta=\tanh^{-1}z/t$  is the space-time rapidity, $g$ depends
on centrality  and defines the multiplicities of produced hadrons,
$U^{\mu}=( \cosh\eta, 0, 0, \sinh\eta)$, $V^{\mu}=(\sinh\eta, 0, 0,
\cosh\eta)$. One can see that $p\cdot U$ and $p\cdot V$ depend on
$\theta=\eta - y$, where $y=\tanh^{-1}p_{L}/p_{0}$, and thus
$f(x,p)$ is longitudinally boost invariant distribution. The
anisotropy of the $f(x,p)$  in momentum plane  is explicitly seen if
Eq. (\ref{10}) is rewritten in the local rest frame, $\eta =0$,
where it is
\begin{eqnarray}
f(x,p)=g
\exp\left(-\sqrt{\frac{p_{T}^2}{\lambda_{\perp}^2}+\frac{p_{L}^2}{\lambda_{\parallel}^2}}\right)\rho(\textbf{r}_{T}).
\label{11}
\end{eqnarray}

First we use the hydrodynamical code \cite{Karp} in its ideal fluid
form (i.e., with zero viscosity coefficients), and perform
calculations with initial conditions defined by  Eq. (\ref{10}). To
make a comparison between isotropic and anisotropic in momentum
space initial distributions, we use different values of $\lambda
\equiv \lambda_{\perp}/\lambda_{\parallel}$: $1$, $0.01$, and $100$
respectively. As for the $\lambda_{\perp}$, we utilize the fixed
value $1.4$ GeV for all calculations. Therefore, $\lambda = 0.01$
corresponds to the large longitudinal pressure, as compared to the
transverse one; $\lambda = 100$ means very small longitudinal
pressure, similar as in original Color Glass Condensate (CGC)
initial conditions (IC) \cite{parametr-2}. The value of $\lambda
\approx 1$ is used, in fact,  in Ref. \cite{parametr-3}. This value
of $\lambda$ corresponds to  CGC-like IC with smeared
$\delta(\eta-y)$ in the gluon CGC Wigner function;  the smearing was
provided to escape contradiction with the quantum uncertainty
principle. Also, we utilize $R_{x}=R_{y}=R=5.33$ fm in Eq.
(\ref{9.1}), which corresponds to the Gaussian approximation of the
initial energy density transverse  profile in central heavy ion
collisions \cite{parametr-3}.  The results for energy densities and
velocities are demonstrated  for the time of thermalization
$\tau=\tau_{\text{th}}=1$ fm/c, when transition to hydrodynamics is
assumed to be fulfilled.

Let us compare the results for the energy densities and transverse
velocities at $\tau=\tau_{\text{th}}$ from the relaxation model
(RM), hydrodynamic model (HM), and free streaming (FS) one. Energy
densities and   four-vector energy flow field are calculated from
energy-momentum tensor:
\begin{eqnarray}
u^{\mu}=\frac{T^{\mu
\nu}u_{\nu}}{T^{\mu\nu}u_{\mu}u_{\nu}}=\frac{T^{\mu
\nu}u_{\nu}}{\epsilon}. \label{field}
\end{eqnarray}
We apply HM and FS models in the following way. For HM we utilize
Eq. (\ref{field}) at $\tau=\tau_{0}$ with $T^{\mu \nu}=T^{\mu
\nu}_{\text{free}}$ to calculate the initial energy density and
four-velocities, then incorporate them in the energy-momentum tensor
in the hydrodynamical form and perform a pure hydrodynamical
evolution until $\tau=\tau_{\text{th}}$, whereas  for FS model
initial energy-momentum tensor at $\tau=\tau_{0}$ fully coincides
with one in RM, and we apply Eq. (\ref{field}) to calculate  the
 energy density and four-velocities in FS model at $\tau=\tau_{\text{th}}$.

Before discussing the results of numerical calculations, let us
perform analytical estimate and comparison of the energy density
evolution in HM and FS models. Since there is no initial transverse
flow, one can calculate energy density $\epsilon({\bf}r_T,\tau)$, at
least in the central part, $r/R \ll 1$, using approximation of
transversely homogeneous system within the times $\tau$ such that
$\tau/R \ll 1$. Then in the hydrodynamic model with EoS of massless
ideal gas, $p=\epsilon/3$, one gets the known result:
\begin{equation}
\epsilon_{\text{HM}}(\tau) \propto \left(\frac{\tau_0}{\tau}\right)^{4/3}.
\label{hm}
\end{equation}
The direct calculation based on Eqs. (\ref{002})  for free streaming regime with
the boost-invariant initial distribution $f({\bf r}_T, {\bf p}_T, \theta)$ (\ref{10}),
 in which the evolution is described by the substitution $\theta \rightarrow \theta(\tau)=\text{arcsinh}(\frac{\tau}{\tau_0}\sinh(\theta))$,
 ${\bf r}_T \rightarrow {\bf r}_T(\tau)={\bf r}_T-\frac{{\bf p}_T}{m_T}(\tau\cosh(\theta)-\sqrt{\tau_0^2+\tau^2\sinh^2(\theta)})$
 \cite{early},  gives
\begin{eqnarray}
\epsilon_{\text{FS}}(\tau)\propto \frac{\tau_0}{\lambda\tau}\frac{\arccos(\frac{\tau_0}{\lambda\tau})}{\sqrt{1-\frac{\tau_0^2}{\lambda^2\tau^2}}} + \frac{\tau_0^2}{\lambda^2\tau^2}.
\label{fs}
\end{eqnarray}
Note that in Eq. (\ref{fs}) we use the equality
$\arccos(x)=i\text{arccosh}(x)$. It is easy to find that in the
non-relativistic analog of boost-invariant free streaming the first
term in Eq. (\ref{fs}) $\propto \frac{\tau_0}{\lambda\tau}$ is
associated with a decrease with time of the transverse energy  due
to a reduction of particle number density  because of the collective
longitudinal expansion (then a gain of the particle number in some
longitudinally small region is less
 than a loss of it). The second term is related to a decrease of
 the longitudinal contribution to the energy density due to
 similar reasons. In the relativistic situation, when one cannot split the particle energy
into the sum of the longitudinal and transverse parts, such an interpretation of Eq. (\ref{fs}) is not so exact.

It is easy to  see that in the limited interval of $\tau$, $(\tau_0,
\tau_{th})$, the different terms in  Eq. (\ref{fs})  can dominate
depending on the anisotropy parameter $\lambda$. At large $\lambda $
the first term dominates, and then at the same initial energy
densities,
$\epsilon_{\text{HM}}(\tau_0)=\epsilon_{\text{FS}}(\tau_0)$, the
final energy density will be larger in the FS case (cf. Eqs.
(\ref{hm}) and (\ref{fs})), while at fairly small $\lambda$ the
result will be the opposite: $\epsilon_{\text{HM}}(\tau_{\text{th}})
> \epsilon_{\text{FS}}(\tau_{\text{th}})$. The simple calculation
with our parameters $\tau_0=0.1$ fm/c, $\tau_{\text{th}}= 1$ fm/c
shows that $\epsilon_{\text{HM}}(\tau_{\text{th}}) \approx
\epsilon_{\text{FS}}(\tau_{\text{th}})$ for  $\lambda \approx 1/3$.

\begin{figure}[H]
     \centering
     \includegraphics[width=0.65\textwidth]{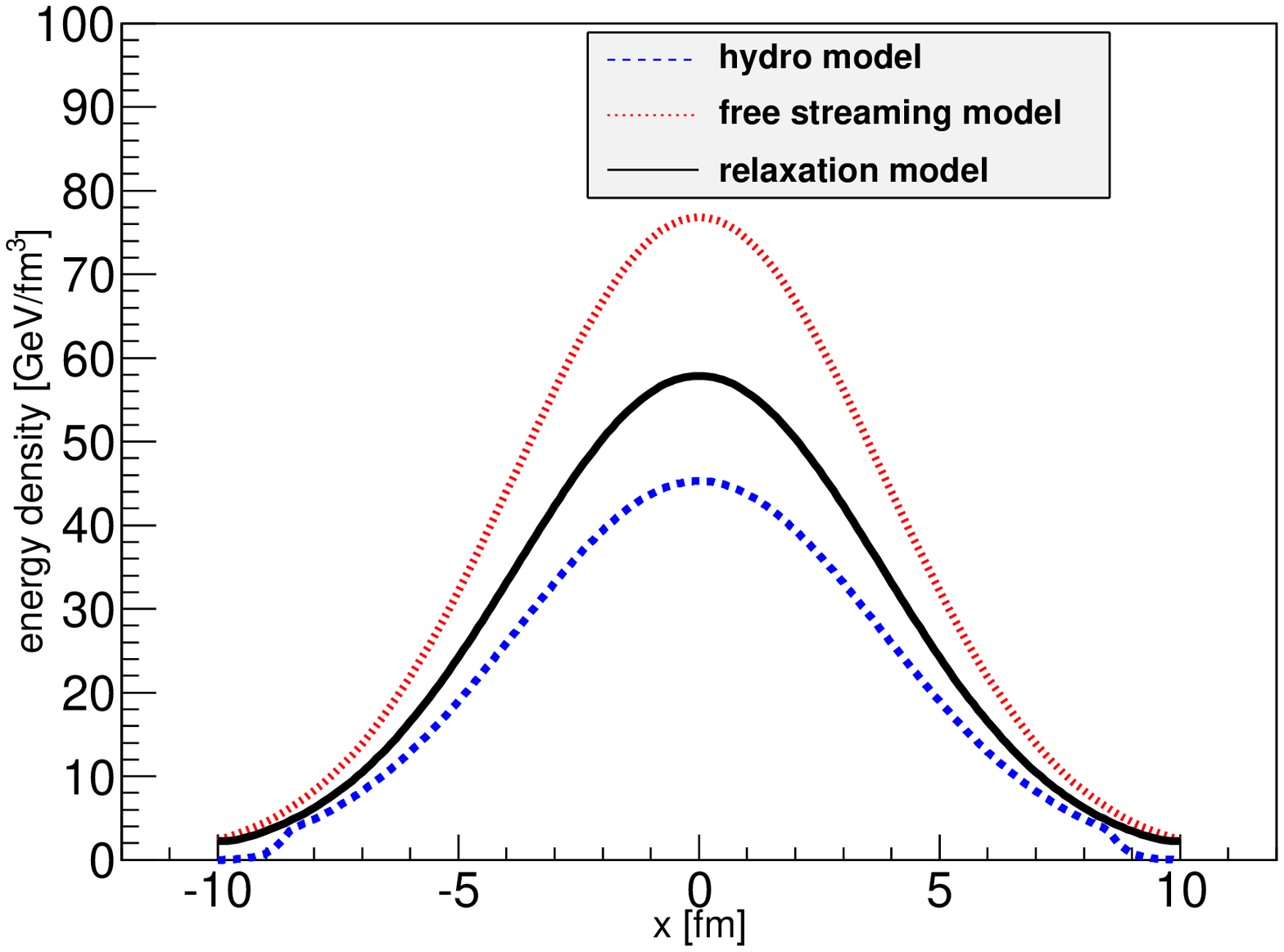}
     \caption{\small
(Color online). The energy density distribution along axis $x$,
($y=0$) in transverse plane for central rapidity slice at
$\tau=\tau_{\text{th}} = 1.0$ fm/c for  the following conditions of
the relaxation evolution: $\tau_{0}= 0.1$ fm/c, the Gaussian initial
transverse energy density profile, $\lambda=1$, EoS: $p=\epsilon/3$,
$\tau_{\text{rel}}=0.5$ fm/c, the target energy-momentum tensor
corresponds to ideal hydrodynamics.} \label{fig:ex_anis1}
\end{figure}

\begin{figure}[H]
     \centering
     \includegraphics[width=0.65\textwidth]{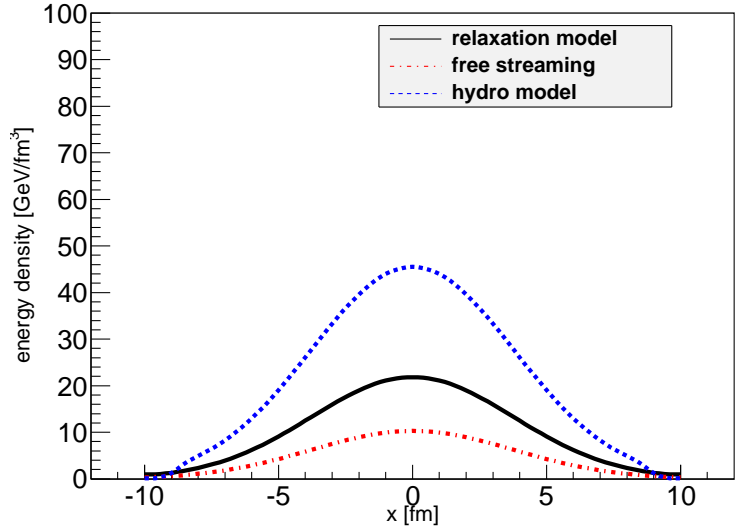}
     \caption{\small
(Color online). The energy density distribution at
$\tau=\tau_{\text{th}} = 1.0$ fm/c under the same conditions as in
Fig. \ref{fig:ex_anis1}, but with large
  initial anisotropy,  $\lambda=0.01$.} \label{fig:ex_anis001}
\end{figure}

\begin{figure}[H]
     \centering
     \includegraphics[width=0.65\textwidth]{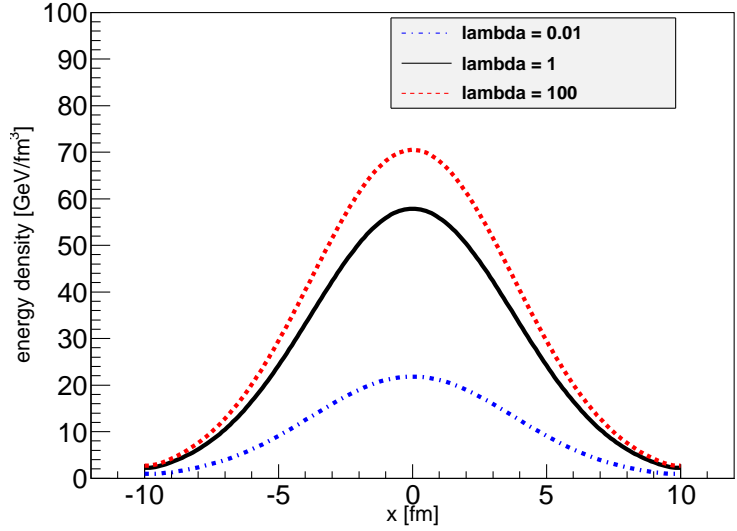}
     \caption{\small
(Color online). The comparison of the results of the relaxation
model for energy density distributions  at $\tau=\tau_{\text{th}} =
1.0$ fm/c for momentum isotropic, $\lambda=1$, and very anisotropic,
$\lambda=0.01$, $\lambda=100$ initial states under the  same other
conditions as in Fig. \ref{fig:ex_anis1}.}
\label{fig:ex_anis_1_001_100}
\end{figure}

In Fig. \ref{fig:ex_anis1} we present the
energy densities  for the relaxation model  in comparison with
the hydrodynamic model   and the free streaming  evolution  for
isotropic initial state, $\lambda =1$, with the equation of state
$p=\epsilon/3$. As one can see from Fig. \ref{fig:ex_anis1}, the final (at the
thermalization time $\tau = \tau_{\text{th}}=1$ fm/c) energy density in RM  is in
between the corresponding results for HM (minimal values) and FS model (maximal
values) cases. These results are expected because the relaxation evolution incorporates both
HM- and FS- regimes, and the anisotropy parameter is chosen to be $\lambda=1>1/3$.
 Note that in any anisotropic case, i.e. $\lambda \neq 1$, which we start to discuss,
 the  initial conditions for hydrodynamics at $\tau=\tau_{0}=0.1$ fm/c are taken in our analysis
 with the same initial energy density as at the anisotropic distribution,
 $\epsilon_{\text{hyd}}=\epsilon$,
but with symmetric pressure $p_{\text{hyd}}= c_0^2\epsilon$, where $c_0^2=1/3$
 everywhere in our analysis except for the specially defined cases.  Figure \ref{fig:ex_anis001}
 corresponds to the anisotropic case with $\lambda=0.01$, when the longitudinal
 ``effective temperature'' $\lambda_{\|}$ is much larger than the transverse one, $\lambda_{\|}> \lambda_{\bot}$.
Again, the final energy densities at  RM- regime reach intermediate values between the corresponding
results for HM- and FS- cases. However, since now  $\lambda<1/3$, the smallest values of the
energy  are reached at FS- regime and the maximal ones at HM- expansion. The detailed analysis
of the RM-, HM- and FS- regimes  in the vicinity of the point $\lambda=1/3$: $\lambda= (0.25,0.45)$
demonstrates the changing of sequence for different regimes at $\tau_{th}$, at the same time RM- energy
density never coincides simultaneously with both HM- and FS- energy densities. Also, a coincidence of the RM
final energy density with either the HM or the FS model one is accompanied by {\it different} pre-thermal
flows in the corresponding pair. In Fig. \ref{fig:ex_anis_1_001_100} we compare the results of
the relaxation model for the initial distributions, which are isotropic, $\lambda = 1$,
and strongly anisotropic, $\lambda= 0.01$ and $\lambda=100$,  in momentum space. The last case
is associated typically with
Color Glass Condensate (CGC) initial conditions \cite{parametr-2}. One can see that at the
 same initial densities at $\tau=0.1$ fm/c,
the maximal energy density at the thermalization time $\tau=1$ fm/c is reached
for the case with the smallest longitudinal  pressure.

\begin{figure}[H]
     \centering
     \includegraphics[width=0.65\textwidth]{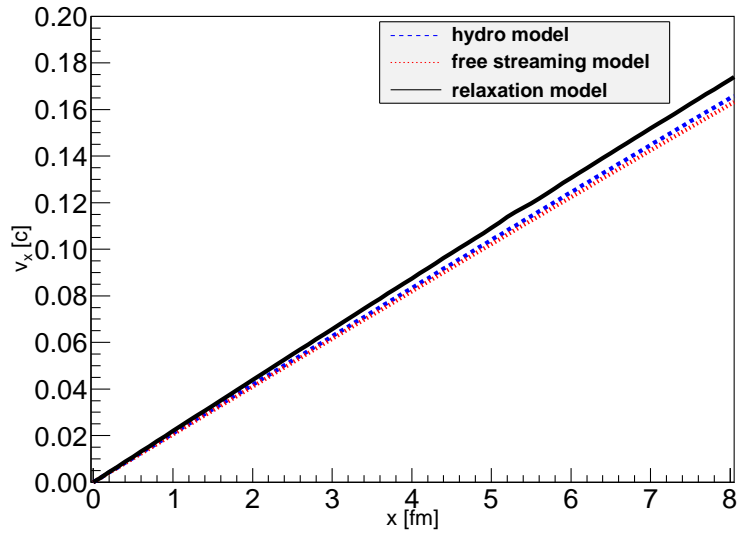}
     \caption{\small
(Color online). The transverse velocity distribution at
$\tau=\tau_{\text{th}} = 1.0$ fm/c and  $\lambda=1$ under the same
conditions as in Fig. \ref{fig:ex_anis1}.} \label{fig:vx_anis1}
\end{figure}

\begin{figure}[H]
     \centering
     \includegraphics[width=0.65\textwidth]{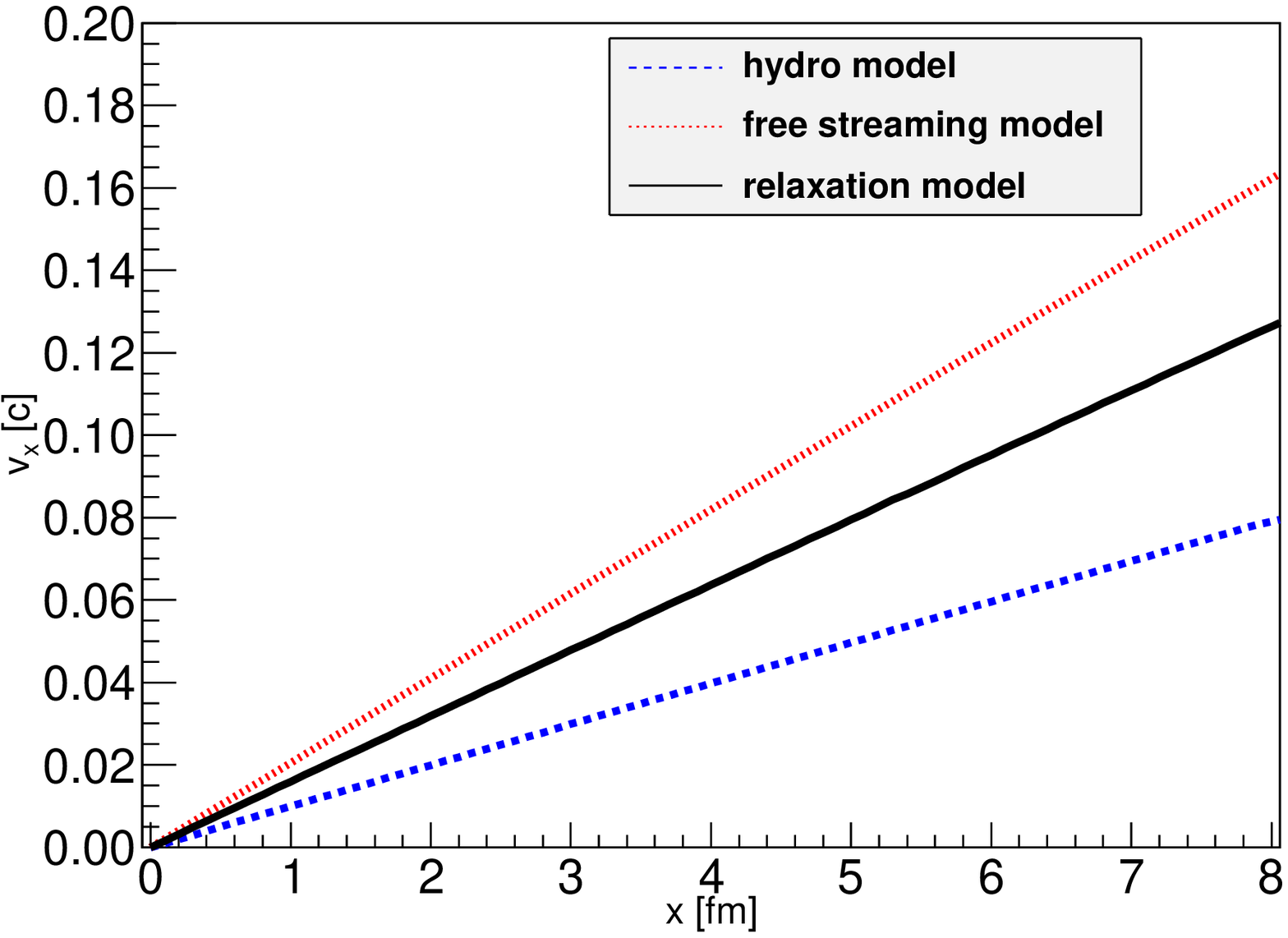}
     \caption{\small
(Color online). The transverse velocity distribution at
$\tau=\tau_{\text{th}} = 1.0$ fm/c under the same conditions as in
Figs. \ref{fig:ex_anis1}, \ref{fig:vx_anis1} but softer EoS:
$p=0.15\epsilon$.} \label{fig:vx_anis_1_015}
\end{figure}

\begin{figure}[H]
     \centering
     \includegraphics[width=0.65\textwidth]{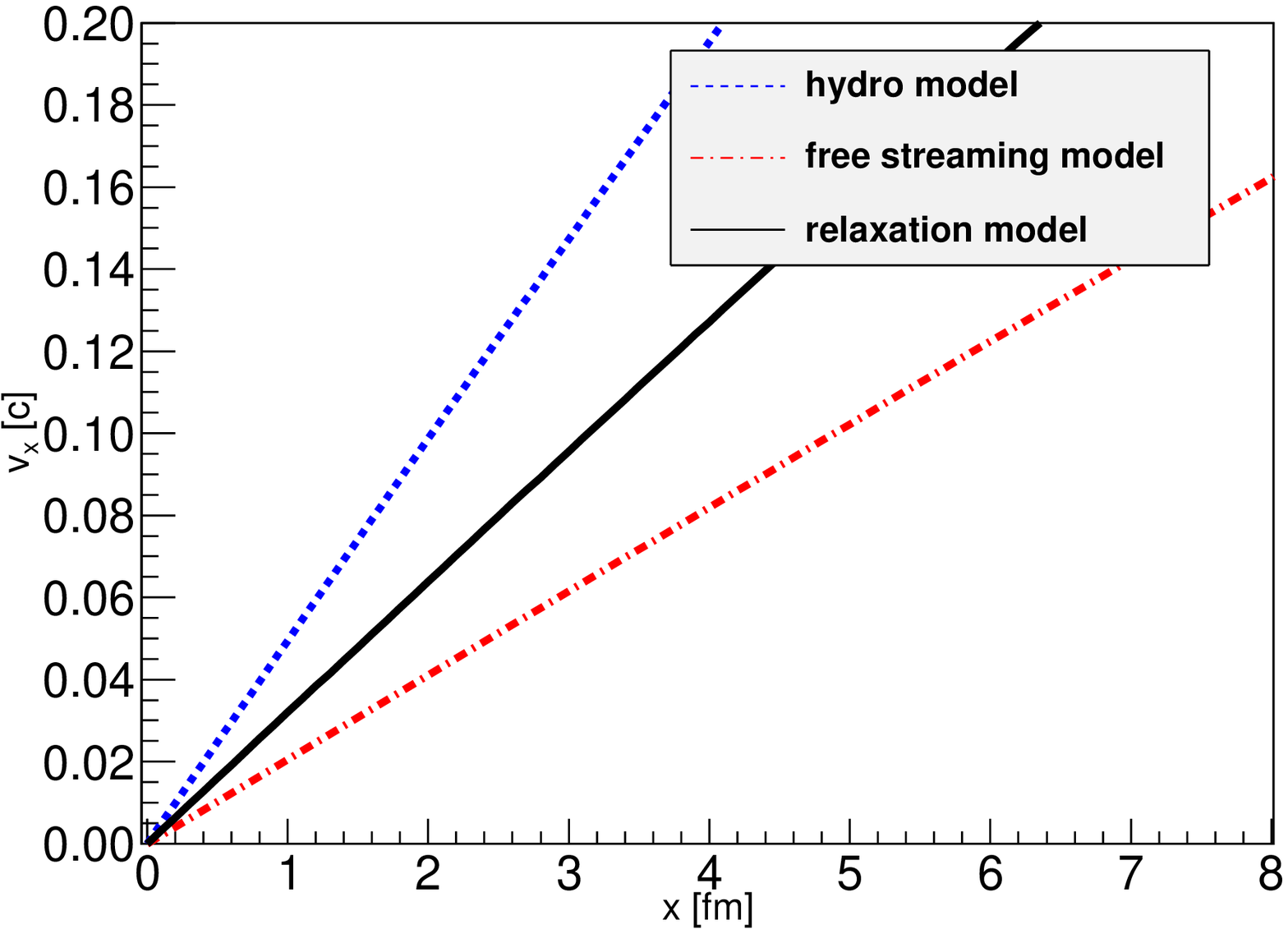}
     \caption{\small
(Color online). The transverse velocity distribution at
$\tau=\tau_{\text{th}} = 1.0$ fm/c under the same conditions as in
Figs. \ref{fig:ex_anis1}, \ref{fig:vx_anis1} but harder EoS:
$p=0.7\epsilon$.} \label{fig:vx07}
\end{figure}

The situation with the transverse collective velocities (Fig.
\ref{fig:vx_anis1}) is not  trivial  even in the isotropic case,
where the energy density value in RM
 is in between the ones in the FS model and HM. The reason is that there are two oppositely
directed factors which act simultaneously to the transverse gradient
of the hydrodynamic pressure that contributes to formation of the
velocity field in the RM-regime. On the one hand, harder equation of
state increases the gradient, but on the other hand, the hard
equation of state could reduce it since more energy loss happens due
to the fact that more work is done in the longitudinal direction by
the system contained in some rapidity interval. We demonstrate such
an interplay in  Figs. \ref{fig:vx_anis1}, \ref{fig:vx_anis_1_015},
and \ref{fig:vx07}; the maximal final velocities are reached at
$p_{\text{hyd}}=\epsilon / 3$ in RM, at $p_{\text{hyd}}=0.15
\epsilon$ in FS, and at $p_{\text{hyd}}=0.7 \epsilon$ in HM.

\begin{figure}[H]
     \centering
     \includegraphics[width=0.65\textwidth]{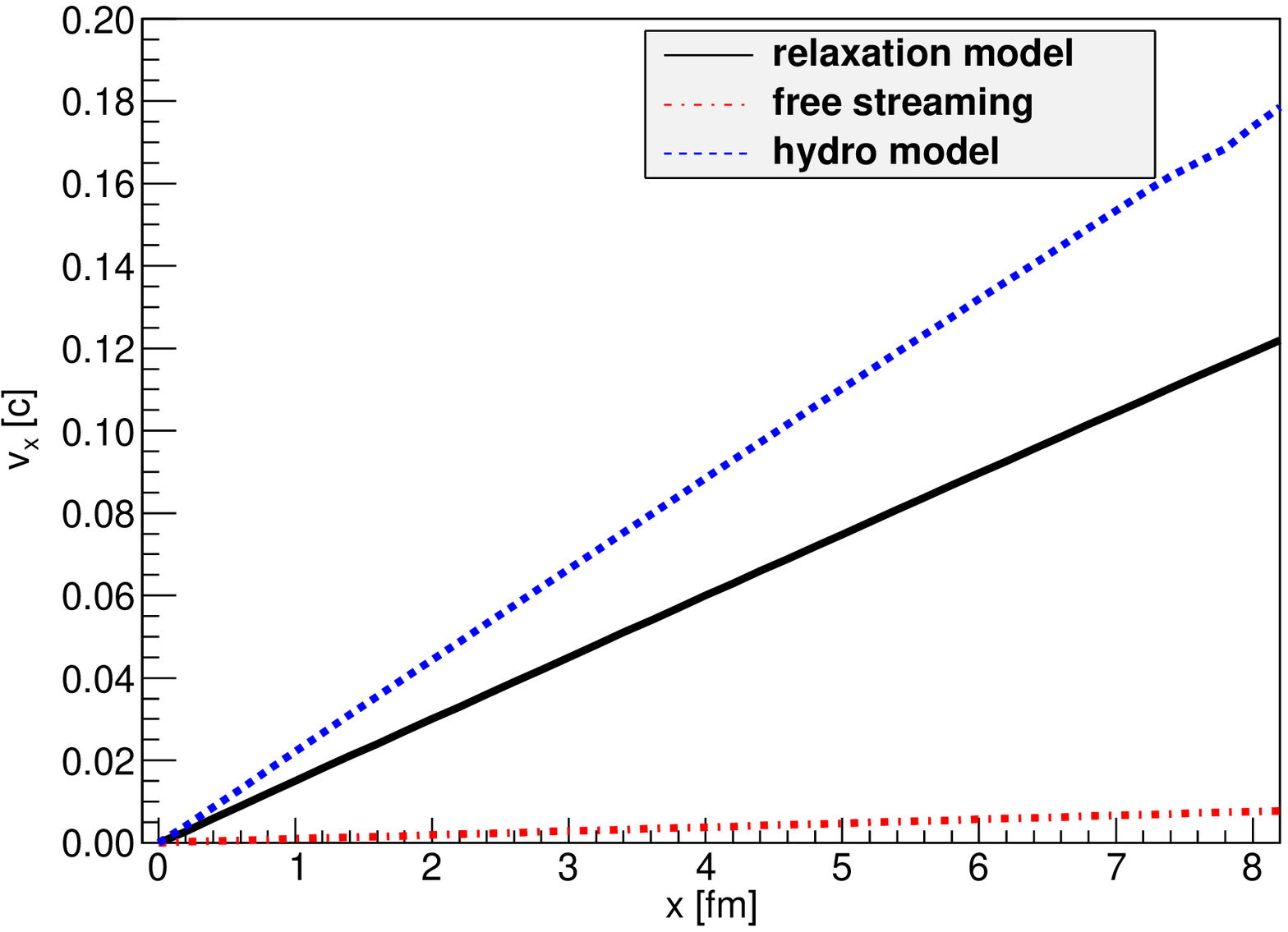}
     \caption{\small
(Color online). The transverse velocity distribution at
$\tau=\tau_{\text{th}} = 1.0$ fm/c under the same conditions as in
Figs. \ref{fig:ex_anis1}, \ref{fig:vx_anis1},
 but with large initial anisotropy,  $\lambda=0.01$.}
\label{fig:vx_anis001}
\end{figure}

\begin{figure}[H]
     \centering
     \includegraphics[width=0.65\textwidth]{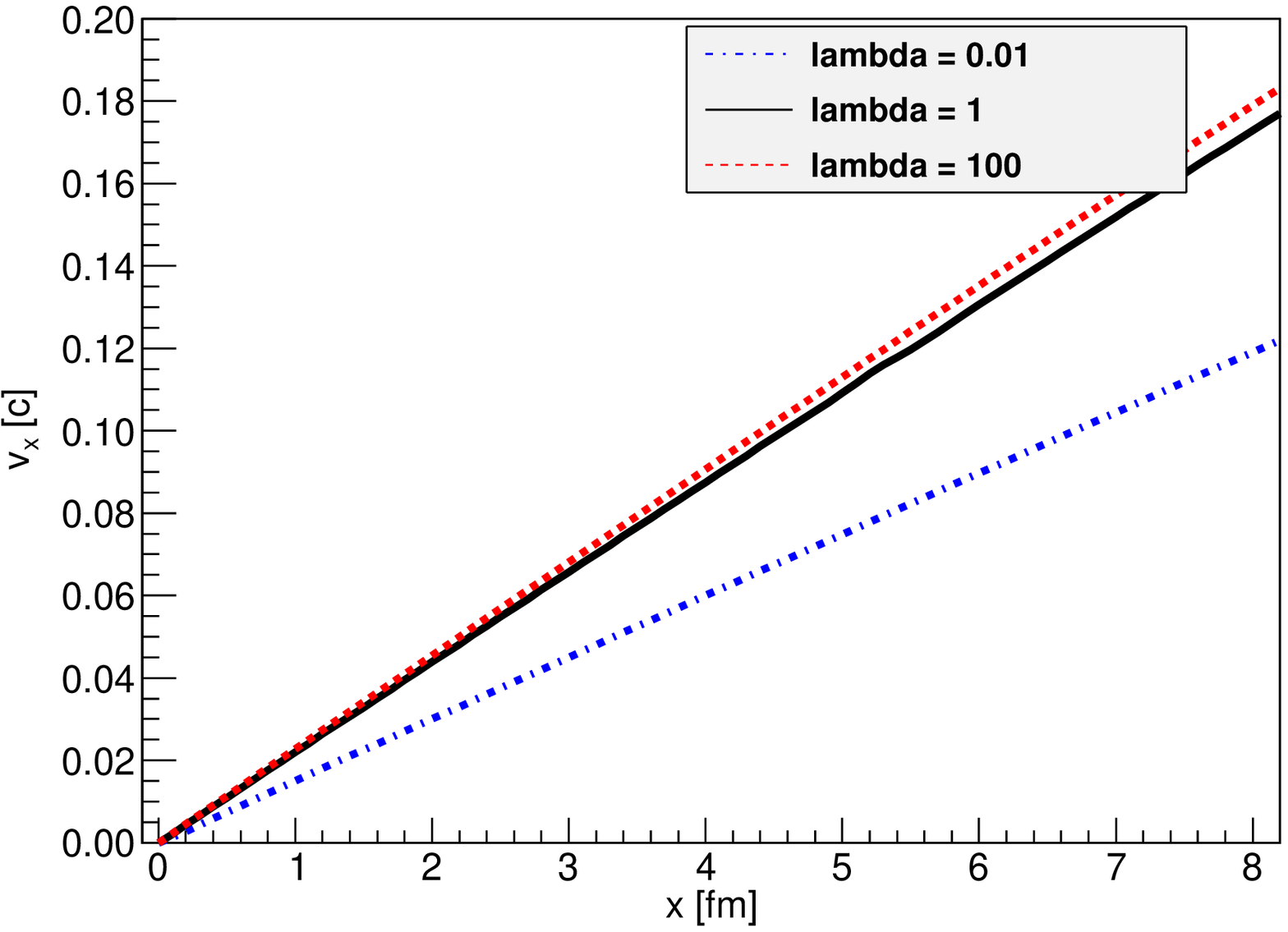}
     \caption{\small
(Color online). The comparison of the results of the relaxation
model for transverse velocity distributions  at
$\tau=\tau_{\text{th}} = 1.0$ fm/c for isotropic, $\lambda=1$ and
very anisotropic, $\lambda=0.01$, $\lambda=100$ initial states under
the  same other conditions as in Fig. \ref{fig:ex_anis1}.}
\label{fig:vx_anis_1_001_100}
\end{figure}

In Fig. \ref{fig:vx_anis001} the transverse velocities for initially
very anisotropic states, $\lambda = 0.01$ are presented. As opposed
to the $\lambda=1$ regime, at small $\lambda<1/3$ the minimal
pre-thermal flow develops in the case of the  FS-regime and a
maximal flow in the HM one. The RM-regime
 leads  to an intermediate result. One can see the influence of the anisotropy on the RM results from Figs.
\ref{fig:ex_anis_1_001_100}, \ref{fig:vx_anis_1_001_100}, where it is
found that the energy densities and pre-thermal collective flows grow when the anisotropy parameter
 $\lambda$ increases. The reason is that the increase of   $\lambda$ results in the decrease of the energy
 density loss rate for  the FS-component during the boost-invariant expansion (see Eq. (\ref{fs}) and discussion there).
As a result, the maximal energy densities and transverse gradient of
the FS-component are larger during the relaxation process when
$\lambda$ is larger. In its turn the transverse gradient of particle
or/and energy densities define the transverse flow at free streaming
process \cite{early}, \cite{hbt-ic}: It grows when the gradient
increases.

\begin{figure}[H]
     \centering
     \includegraphics[width=0.65\textwidth]{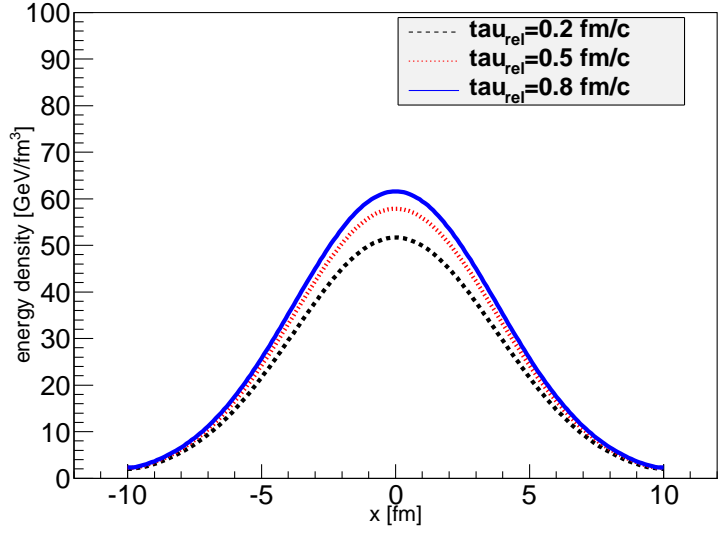}
     \caption{\small
(Color online). The comparison of the  energy density distributions
at different $\tau_{\text{rel}}=0.2, 0.5, 0.8$ fm/c under the  same
other conditions as in Fig. \ref{fig:ex_anis1}.}
\label{fig:ex_anis_1_tau_rel}
\end{figure}

\begin{figure}[H]
     \centering
     \includegraphics[width=0.65\textwidth]{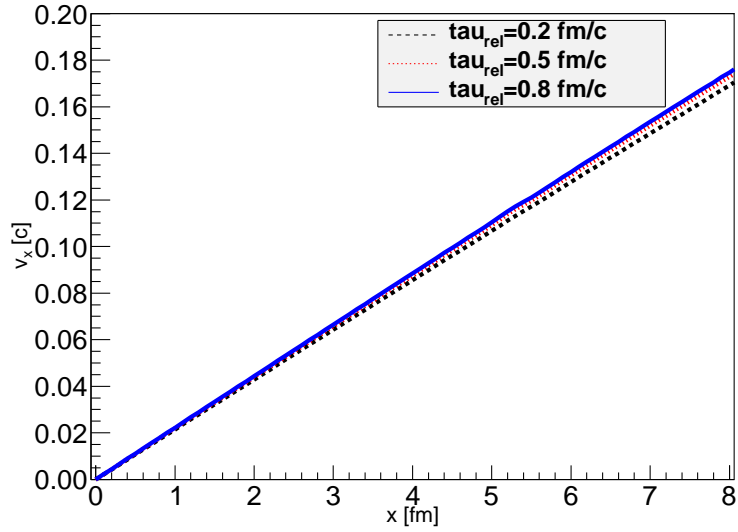}
     \caption{\small
(Color online). The comparison of the transverse velocity
distributions at different $\tau_{\text{rel}}=0.2, 0.5, 0.8$ fm/c
under the  same other conditions as in Figs. \ref{fig:ex_anis1},
\ref{fig:vx_anis1}.} \label{fig:vx_anis_1_tau_rel}
\end{figure}

To study how the rate of transition to hydrodynamics affects the
energy densities and transverse velocities at the starting time of
the hydrodynamic stage, we perform the simulations for different
values of $\tau_{\text{rel}}$. The results are demonstrated in Figs.
\ref{fig:ex_anis_1_tau_rel}, \ref{fig:vx_anis_1_tau_rel} for
$\lambda =1$ where one can see that the final transverse velocities
are almost the same despite the different slopes of the transition,
whereas  the energy densities are larger if the rate of transition
to hydrodynamics is smaller ($\tau_{\text{rel}}$ is larger). Such a
behavior in the initially symmetric case is due to the free
streaming regime, which dominates for longer time at larger
$\tau_{\text{rel}}$. This also means that at $\lambda=1>1/3$ the
energy density will be larger, cf. Eqs. (\ref{hm}) and (\ref{fs}).
At the same time, by comparing the energy densities and transverse
velocities, calculated in RM with $\tau_{\text{rel}}=0.8$ fm/c, that
corresponds to rather abrupt transition to hydrodynamics near the
end of the pre-equilibrium stage, with the ones calculated in FS,
see Figs. \ref{fig:ex_anis1}, \ref{fig:vx_anis1}, one can find
noticeable differences in the resulting energy density, which is
smaller in RM than at the  free streaming regime with $\lambda =1$.
As a matter of fact, one cannot get the free streaming
 regime in the relaxation model.

It is important to emphasize now that both FS- and HM- regimes of expansions are not real
 limiting cases for the system
evolution from an initial non-equilibrated (NEQ) state to a final
equilibrated (EQ) one. At the same initial energy density profile,
for any allowed parameters: $\tau_{0}, \tau_{\text{th}}$ and
$\tau_{\text{rel}}$, such ``limiting''  energy-momentum tensors
cannot be reached. Formally, basic boundary equalities (\ref{3})
prevent such a possibility. The physical reason is that the
structures of the energy-momentum tensor are different for EQ- and
NEQ- states. As we have discussed above, both the final energy
density and transverse velocity profile in relaxation model do not
coincide simultaneously with analogous values reached at FS- or HM-
regimes, for any value of the anisotropy parameter $\lambda$. In the
particular case of isotropic initial state ($\lambda =1$), the
energy-momentum tensor in the free streaming evolution acquires
specific (non-viscous)  non-equilibrium structure with the energy
density higher than in the case of any continuous transition to
hydrodynamics demonstrated at Fig. \ref{fig:ex_anis_1_tau_rel}. The
reason is that one cannot simply ignore the non-diagonal terms in
the energy-momentum tensor that are developing at free streaming
evolution \cite{hbt-ic}, and diagonalize the tensor ``suddenly'' at,
say, $\tau = \tau_{\text{th}} = 1$ fm/c by using  Landau matching
conditions, because then the energy-momentum conservation is
violated. Thereby, the continuous interpolation between arbitrarily
anisotropic initial state and hydrodynamical regime  at some later
time, provided by the relaxation model, does not imply ``continuous
interpolation'' between the two ``limiting'' types of the matter
evolutions: free streaming and hydrodynamical.

\begin{figure}[H]
     \centering
     \includegraphics[width=0.65\textwidth]{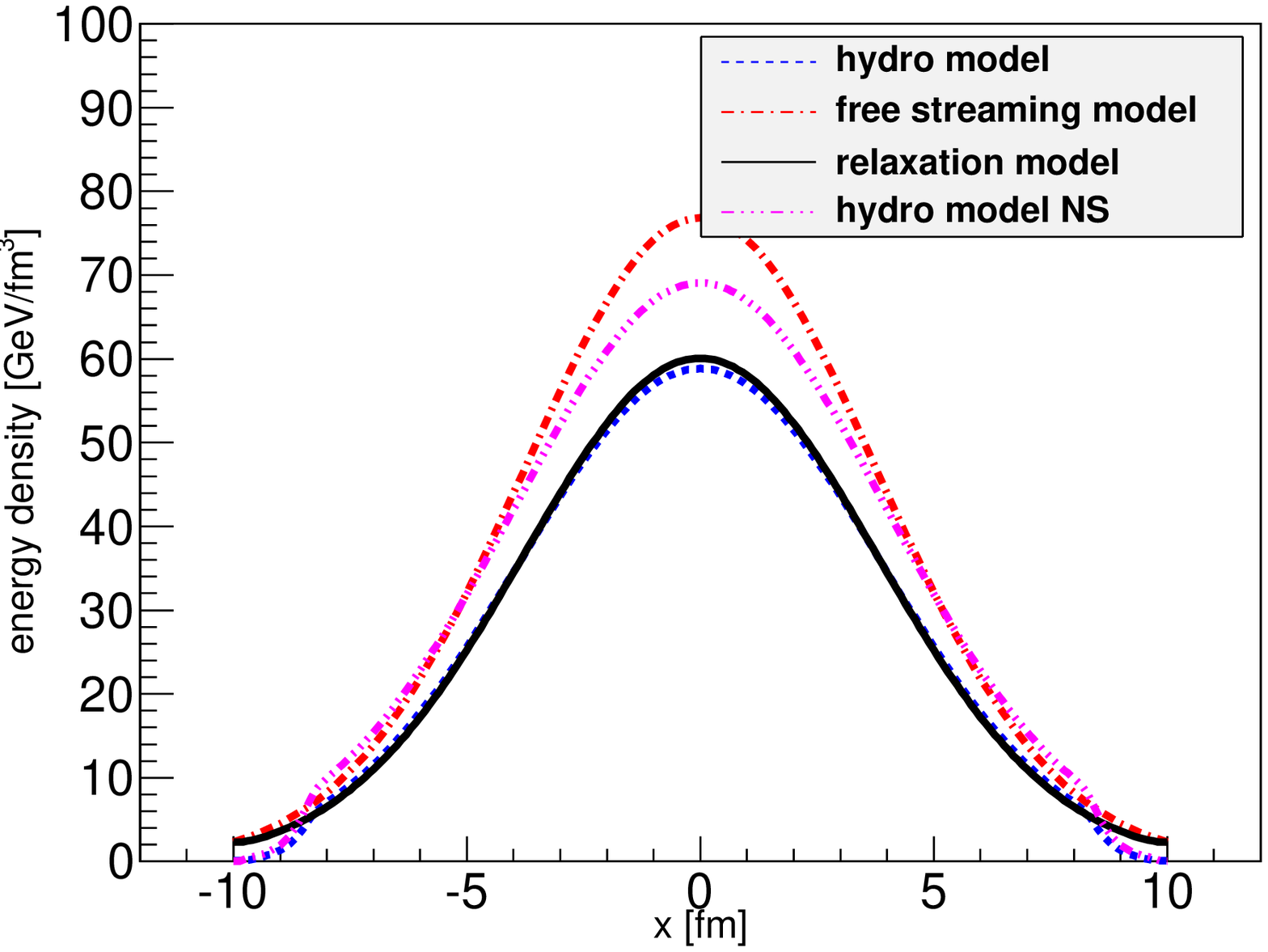}
     \caption{\small
(Color online). The energy density distribution along axis $x$,
($y=0$) in transverse plane for central rapidity slice at
$\tau=\tau_{\text{th}} = 1.0$ fm/c for  the following conditions of
the relaxation evolution: $\tau_{0}= 0.1$ fm/c, the Gaussian initial
transverse energy density profile, $\lambda=1$, EoS: $p=\epsilon/3$,
$\tau_{\text{rel}}=0.5$ fm/c, the target energy-momentum tensor
corresponds to viscous hydrodynamics with $\eta/s$=0.1.}
\label{fig:ex_0.1}
\end{figure}

\begin{figure}[H]
     \centering
     \includegraphics[width=0.65\textwidth]{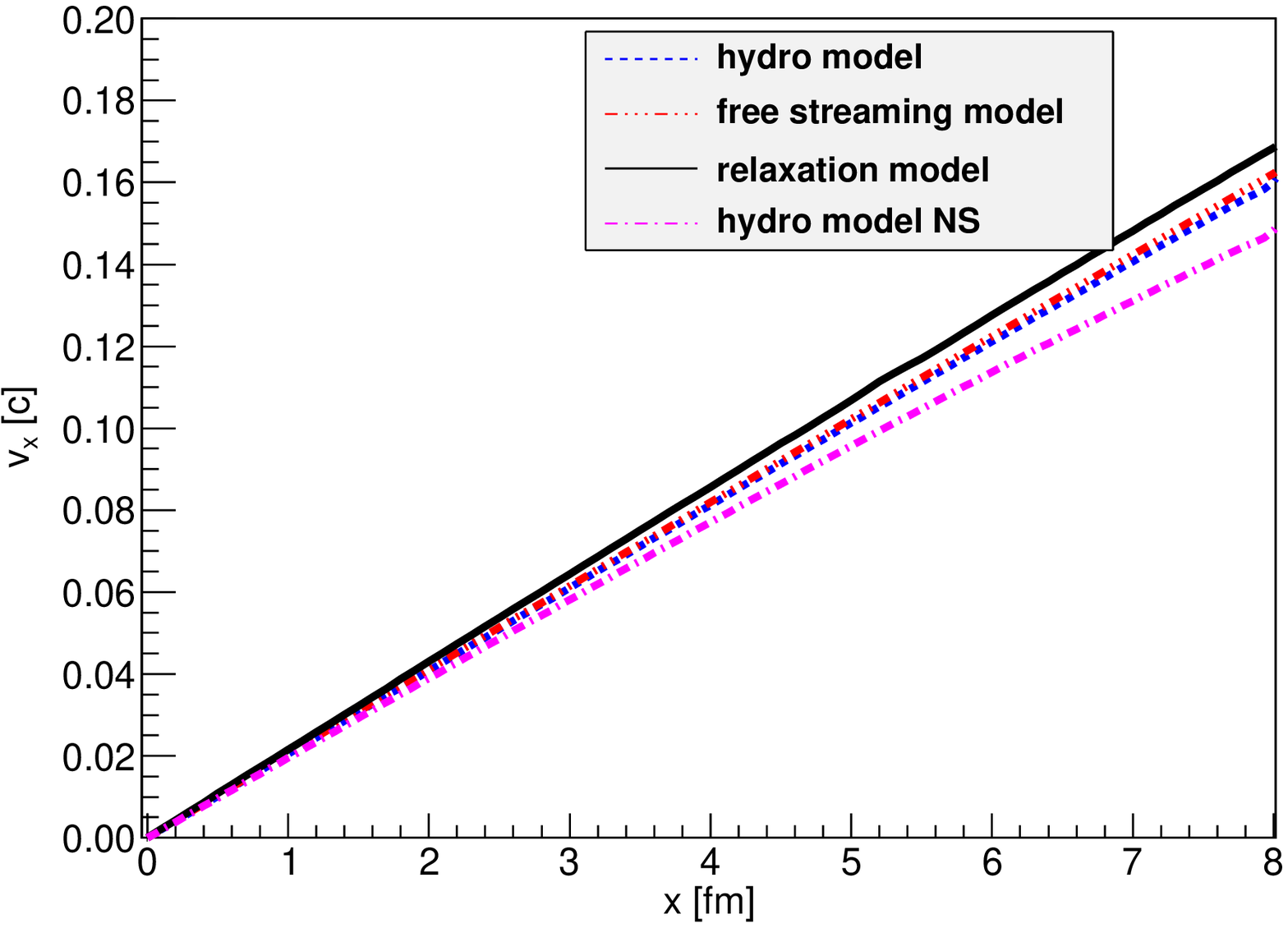}
     \caption{\small
(Color online). The transverse velocity distribution at
$\tau=\tau_{\text{th}} = 1.0$ fm/c under the same conditions as in
Fig. \ref{fig:ex_0.1}.} \label{fig:vx_0.1}
\end{figure}

Another important point, which we  outline here, is the
initialization of viscous hydrodynamics by means of the relaxation
model. The relaxation model equations (\ref{8}) include the
Israel-Stewart viscous hydrodynamical formalism, and the
corresponding viscous hydrodynamic code \cite{Karp} is modified to
solve the relaxation model equations. We keep bulk pressure $\Pi=0$
everywhere during the evolution, by default use zero initial values
for the shear stress tensor, $\pi^{\mu\nu}(\tau_{0})=0$, and use
anzatz for the relaxation time of the shear stress tensor:
$\tau_\pi=5\eta/(sT)$. In the viscous case one also needs to specify
the temperature and entropy density from the equation of state.
Therefore for the viscous case we always use the EoS for a
relativistic massless gas of quarks and gluons, $p=\epsilon/3$,
which results in the following relation for energy density:
\begin{eqnarray}
 \epsilon=\left(\frac{7}{4} g_l  n_f +
g_g\right)\frac{\pi^2}{30}T^4, \label{11-0}
\end{eqnarray}
where $g_l=6$ ang $g_q=16$ are quark and gluon degeneracy factors,
respectively, and we set the effective number of quark degrees of
freedom $n_f=2.5$. Equation (\ref{11-0}) as well assumes that the
chemical potentials are zero. The entropy density is then extracted
using the  thermodynamical relation $\epsilon+p=Ts$. In the limiting
Navier-Stokes case the fixed ratio of shear viscosity to entropy
density $\eta/s=0.1$ is used. Also, in this case we perform
calculations with initially isotropic momentum space distribution,
$\lambda = 1$. The results of RM are compared with the results of HM
with initial shear viscous tensor $\pi^{\mu\nu}(\tau_{0})=0$, and
with results of HM with initial shear viscous tensor equals to its
limiting Navier-Stokes (NS) form, $\pi^{\mu\nu}
(\tau_{0})=\pi^{\mu\nu}_{NS}$. The results are shown in Figs.
\ref{fig:ex_0.1} and \ref{fig:vx_0.1}, where the free streaming
model calculations are added for comparison. The results of the
relaxation model with zero viscous initial conditions are
qualitatively similar to the results obtained in the relaxation
model with the ideal fluid, for example, one can see from these
figures that the relaxation model results in lower energy density in
the center of the system and in higher transverse velocities as
compared to the free streaming model. In the case of non-zero
initial viscosity contribution at $\tau = \tau_0$ (NS limit), the
final energy density in such a model is larger than in the
relaxation model.

Finally  we demonstrate that the viscous relaxation  model, unlike
``anisotropic hydrodynamics'', can be used not only for smooth
initial energy density profiles, but also for bumpy anisotropic
initial distributions at $\tau = \tau_0$.  The set of such bumpy
individual samples is used in  event-by-event hydrodynamic analysis.
To avoid misunderstanding, note that our aim here is not analysis
which of the models: MC-Glauber, MC-KLN or IP-Glasma is more
realistic for a data description (for this aim one needs to utilize
a large number of  individual samples and describe hadron momentum
spectra and their $v_n$-coefficients), but only the demonstration of
capabilities of the relaxation model. Thus, we pick randomly one
particular (fluctuating) event corresponding to 20-30\% central
Pb+Pb collision at LHC energy generated with the GLISSANDO
generator \cite{generator}. The generator calculates the spatial
distribution of sources, which fluctuates according to the
statistical nature of the distributions of nucleons in the colliding
nuclei, according to the Monte-Carlo Glauber model. The transverse
coordinates of the sources correspond to either the centers of the
wounded nucleons or the centers of  the binary collisions of
nucleons in colliding nuclei. The corresponding relative deposited
strength (RDS) of each source is $(1-\alpha)/2$ if the source comes
from a wounded nucleon, or $\alpha$ if it comes from a binary
collision. We use the default value $\alpha=0.14$.

\begin{figure}[H]
\centering
\includegraphics[width=0.65\textwidth]{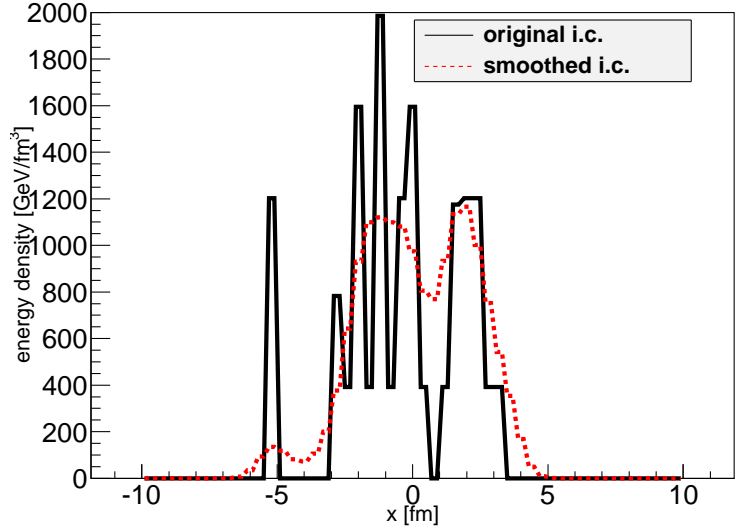}
\caption{\small (Color online). The initial transverse distribution
at $\tau =\tau_{0}= 0.1$ fm/c of the  energy density along $x$-axis
($y=0$) for the random single event produced by the GLISSANDO
generator within Monte Carlo Glauber model. Smoothed distribution is
shown by the dashed line.} \label{fig:id}
\end{figure}

\begin{figure}[H]
\centering
\includegraphics[width=0.65\textwidth]{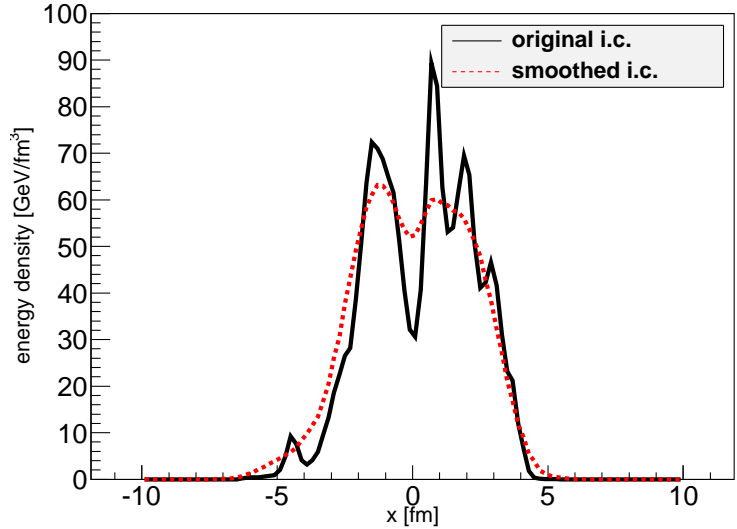}
\caption{\small (Color online). The final energy density
distribution along axis $x$ ($y=0$) in transverse plane for central
rapidity slice at $\tau=\tau_{\text{th}} = 1.0$ fm/c for GLISSANDO
original and smeared initial energy density profiles shown in Fig.
\ref{fig:id}.
 The following conditions of the
relaxation evolution are used: $\tau_{0}= 0.1$ fm/c, $\lambda=1$,
EoS: $p=\epsilon/3$, $\tau_{\text{rel}}=0.5$ fm/c, the target
energy-momentum tensor corresponds to viscous hydrodynamics with
$\eta/s=0.25$.} \label{fig:ex_025}
\end{figure}

\begin{figure}[H]
\centering
\includegraphics[width=0.65\textwidth]{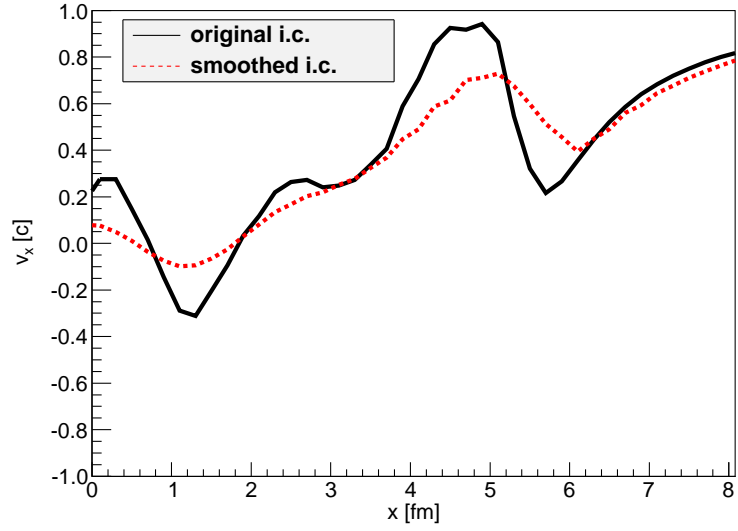}
\caption{\small (Color online). The transverse velocity distribution
for GLISSANDO original and smeared initial energy density profiles
at $\tau=\tau_{\text{th}} = 1.0$ fm/c under the same conditions as
in Fig. \ref{fig:ex_025}.} \label{fig:vx_025}
\end{figure}

One needs initial conditions, which are averaged within each cell,
as a numerical input for the hydrodynamic or relaxation model. In
order to get them, one can use one of standard outputs of GLISSANDO,
which is a $2D$ histogram filled with RDS of all sources in the
transverse plane. The default value of the bin size of the histogram
is $s_{\text{GL}}=0.4$ fm. Formally it is similar to the
coarse-graining procedure which is the basis of any macroscopic
description (say, hydrodynamic) of a microscopic system. In fact,
each hydrodynamic initial condition corresponds to many microscopic
initial states with almost the same densities associated with the
selected scale.\footnote{In fact, one has to deal with sub-ensemble
of GLISSANDO events having close initial  density distributions.
Then hydrodynamics describes the behavior of the mean (macroscopic)
quantities in such a sub-ensemble.} So, if one  wants to associate
the single GLISSANDO  event with further hydrodynamic evolution, the
corresponding energy density distribution cannot be very
inhomogeneous. Also, it is known that the viscosity parameters are
related directly to the coarse-grained  scale \cite{scale}. At too
small scales, the inhomogeneity of the medium can be so large (tends
to the point-like one in the limiting case) that viscous
hydrodynamics loses its applicability. To address this question we
compare the results obtained for the (non-smeared) initial
distribution described above with the results calculated for a
smeared initial distribution. For the latter, we increase the bin
size of the histogram from the default one, $s_{\text{GL}}=0.4$ fm,
to $s_{\text{GL}} =0.7$ fm. To get the smeared initial distribution,
we also distribute the energy from every individual cell to the
transverse area centered around it, according to a Gaussian profile
with a radius $\sigma=0.5$ fm. The non-smeared energy density
profile, produced by GLISSANDO generator  in the randomly selected
single event, is demonstrated in Fig. \ref{fig:id} in comparison
with the smeared one. Note that both bumpy initial distributions,
original and smeared ones, are normalized to the same mean value
$\bar{\epsilon}_0=1000$ GeV/fm$^3$ of the energy density  averaged
in transverse plane within the central square with side $4$ fm. We
use  both, non-smeared and smeared profiles, instead of the Gaussian
one (\ref{9.1})  in Eq. (\ref{10}). Then, using the original and
smeared $r_{T}$-profiles, the corresponding initial phase-space
densities $f(x,p)$, as well as the initial energy-momentum tensors,
can be defined on an event-by-event basis. The results of the
relaxation model for the randomly selected single event with
corresponding initial  energy density distributions in Fig.
\ref{fig:id} are presented in Figs. \ref{fig:ex_025} and
\ref{fig:vx_025} for the case of relatively large viscosity
parameter, $\eta/s=0.25$,  and $\lambda=1$. One can see the
different level of inhomogeneity of initial conditions for
hydrodynamics at $\tau_{\text{th}}=$1 fm/c with smeared and
non-smeared initial states at $\tau=\tau_0$.

Simulations of $A+A$ collisions require initialization of the
relaxation transport model for many such fluctuating initial states.
The statistically relevant number of such events is of the order of
event number $N$ that saturates the average value of initial
conditions at $\tau_0$. It is worth noting that for the GLISSANDO
event generator $N \approx 1000$. Calculation in  the relaxation
transport model of  a single event, that includes also evaluation of
the source term $T^{\mu \nu}_{\text{free}}(x)\partial_{;\mu} {\cal
P}(\tau)$ in Eq. (\ref{8}),  takes about $4$ hours at one processor
core. So, with parallel calculations ($100$ processors) the total
time for event-by-event analysis of relativistic heavy ion
collisions takes about one or two days.

The dependence of the hadron spectra and $v_n$ on the
coarse-graining scale as well as connection between the scale and
viscous parameters is the separate important topic which is beyond
the scope of the present paper.

\section{Summary}

A reliable analysis of the properties of quark-gluon plasma and
initial state of matter formed in relativistic heavy ion collisions
requires  the knowledge  of the pre-thermal dynamics of the
collisions that leads to thermalization of the system and its
further hydrodynamic evolution.  However, until now there has been
no fully satisfactory framework to address dynamical aspects of
isotropization and thermalization in nucleus-nucleus collisions.
Hence a consistent match of a non-equilibrium initial state of
matter with hydrodynamic approximation in such collisions remains an
open question.

We have presented the results for hydrodynamical initial conditions
obtained with the simulations of pre-equilibrium relaxation dynamics
in the energy-momentum transport phenomenological model that was
proposed in Ref. \cite{preth}. Unlike the anisotropic hydrodynamics
 approach \cite{aniz}, where the artificial
concept of ``anisotropic equilibrium'' based on $0+1$ dimensional
kinetics for specific class of anisotropy is utilized instead of
Gibbs relations to close the system of evolutionary equations, this
model does not have any additional assumptions and therefore can be
applied for systems which are arbitrary anisotropic in momentum
space as well as inhomogeneous in transverse plane. The latter is
particularly important for the event-by-event hydrodynamical
modeling of relativistic heavy ion collisions. We have calculated
the initial conditions for the hydrodynamical evolution using
initial states which can be initially isotropic or anisotropic in
momentum space. The dependence of the target thermal state on the
rates of conversion to hydrodynamical regime and different equations
of state is presented as well. It allows us to study the influence
of peculiarities of early initial state as well as pre-equilibrium
dynamics on the energy densities and collective transverse
velocities at the starting time of the hydrodynamical evolution.

In particular it is found that, with the same initial energy density, both final energy densities
 and pre-thermal transverse collective flows increase when the anisotropy parameter -
 the ratio of transverse pressure to longitudinal one,
  increases. Therefore the highest hydrodynamical energy densities and
transverse velocities at  thermalization time are reached at initial zero longitudinal pressure that
 corresponds the CGC-like initial state.  Also, we found that for any relaxation
 time and initial momentum anisotropy, both the final energy density and transverse
 velocity profile in the relaxation model never coincide simultaneously with analogous
 values reached in the hydrodynamic model or at the free streaming regime. Therefore,
continuous relaxation dynamics from an initially non-equilibrium
state to an (almost) equilibrium one can not be properly
approximated by  the free streaming or hydrodynamic regime. The
commonly used prescription of sudden thermalization of the  free
streaming pre-thermal evolution results in discontinuity in the
energy-momentum tensor, which for free streaming has specific
(non-viscous) non-equilibrium structure. This results in a breakdown
of the energy and momentum conservation laws.

The peculiarities of the pre-thermal evolution also depend on the
equation of state for the hydrodynamic component of the system. The
two oppositely directed factors act simultaneously to the transverse
gradient of the hydrodynamic pressure that contributes to formation
of the transverse velocity field at relaxation evolution. On the one
hand, harder equation of state increases the gradient, but on the
other hand, the hard equation of state could reduce it since more
energy loss happens due to more work is done in longitudinal
direction by the system contained in some rapidity interval. As a
result, the maximal transverse velocities are reached   for
isotropic initial state in the following cases: for a soft equation
of state (EoS) at the free streaming evolution, for an ultra-hard
EoS at the pure hydrodynamic expansion, and for an intermediate EoS
for the relaxation evolution.

The developed relaxation model is also applied  for the
situations when the pre-thermal system relaxes to a
close-to-equilibrium state described by viscous hydrodynamics. It is
demonstrated  that  the viscous relaxation model can
be utilized  even with rather bumpy initial states, which allows one to use the model as
a component of hydrodynamical event-by-event analysis.

The physically clear explanations of the results allow one to
conjecture that although  the presented results are model-dependent,
it is plausible to assume that they reproduce  general properties of
the pre-equilibrium dynamics for anisotropic initial momentum
distributions.

\begin{acknowledgments}
S.A. thanks  J. Berges for discussions. S.A. gratefully acknowledges
support  by the DAAD (German Academic Exchange Service). Iu.K.
acknowledges the financial support by Helmholtz International Center
for FAIR and Hessian LOEWE initiative. The research was carried out
within the scope of the EUREA: European Ultra Relativistic Energies
Agreement (European Research Group: "Heavy Ions at Ultrarelativistic
Energies") and is supported by the National Academy of Sciences of
Ukraine, Agreement F6-2015.
\end{acknowledgments}

\pagebreak
\pagebreak

\pagebreak
\pagebreak
\pagebreak

\pagebreak

\pagebreak

\end{document}